\documentclass[aps,pra,reprint,nofootinbib,showkeys,floatfix]{revtex4-2}

\usepackage{amsmath}
\usepackage{graphicx}
\usepackage[colorlinks=true,citecolor=blue,urlcolor=blue]{hyperref}
\usepackage{threeparttable}

\begin{document}

\title{Ferroelectricity and antiferroelectricity in the BaS--PbS system \\
with the rocksalt structure}

\author{Alexander I. Lebedev}
\affiliation{Physics Department, Lomonosov Moscow State University, 119899 Moscow, Russia}


\begin{abstract}
The ferroelectric instability in superstructures, superlattices, quantum wires,
and disordered solid solutions in the BaS--PbS system with the NaCl structure
has been discovered and investigated using first-principles calculations within
the density functional theory. The emergence of ferroelectricity in these structures
is associated with the instability of TO phonons in linear --Pb--S--Pb--S-- chains,
which arises as a result of stretching of the structures upon the introduction
of large barium atoms. Additionally, it has been discovered that, alongside
ferroelectric phases, the structures also exhibit stable, competing
antiferroelectric phases and those with a mixed ferroelectric--antiferroelectric
ordering (ferroelectrically polarized one-dimensional Pb--S chains arranged in an ordered
or disordered manner in the perpendicular direction). These phases often become the
ground state of the studied systems. The closeness of the energies of the ferroelectric,
antiferroelectric, and mixed states indicates the emergence of a multi-minimum
potential with an infinite number of wells separated by potential barriers in the
configuration space. This suggests a possible emergence of nonergodicity in the
structures at low temperatures.

\medskip

\noindent
The accepted manuscript of the paper published in \\
\href{http://dx.doi.org/10.1016/j.jpcs.2026.113887}{Journal of Physics and Chemistry of Solids {\bf 218}, 113887 (2026)}. \\
It is distributed under the CC-BY-NC-ND license.
\end{abstract}


\maketitle


\section{Introduction}

A search for new ferroelectrics with a structure different from the perovskite one
is of interest for expanding our understanding of the phenomenon of ferroelectricity.
In this work, we discover that the BaS--PbS system, which crystallizes in the
NaCl structure (space group $Fm{\bar 3}m$), exhibits ferroelectric properties.
Until now, among the crystals with this structure,
the ferroelectric properties have only been observed in SnTe, GeTe, and in a
number of solid solutions of A$^4$B$^6$ semiconductors~\cite{JChemPhys.44.3323, 
SolidStateCommun.8.1275,JPhysSocJap.38.443,PhysRevLett.37.772,ActaMetall.11.447, 
PhysRevB.18.4920,InfraredPhys.18.877,FizTverdTela.25.3571,JETPLett.40.998,FizTverdTela.28.3610,
JETPLett.46.536,FizTverdTela.32.1780,Ferroelectrics.143.91,JAlloysComp.203.51}.

Unlike the well-studied SrS--PbS system, the BaS--PbS system remains poorly understood.
Only two theoretical papers have been published on the properties of solid
solutions in this system~\cite{JAlloysComp.694.1348,ComputCondensMatter.21.e00398},
along with a report on the preparation of Pb$_{1-x}$Ba$_x$S nanoparticles with a
Ba content of up to 5\% and an orthorhombic structure~\cite{IntJAdvSciEng.3.308}.
In this paper, we demonstrate that the BaS--PbS system exhibits a new, previously
unknown property---ferroelectricity---which is absent in the SrS--PbS system. The unexpectedness
of its appearance in the considered system is due to the fact that it appears in a
system where neither of the initial components (BaS and PbS) are ferroelectrics.

\section{Research objects and calculation details}
\label{sec2}

\begin{figure}
\centering
\includegraphics{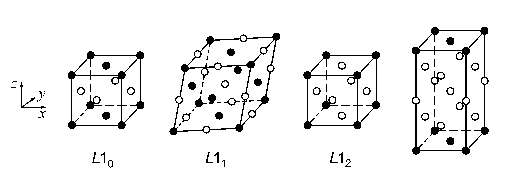}
\caption{Unit cells of three simplest superstructures, $L1_0$, $L1_1$,
and $L1_2$, based on the FCC lattice, and the unit cell of the (BaS)$_3$/(PbS)$_1$
superlattice. The filled and empty dots indicate Pb and Ba atoms, respectively.
Sulfur atoms are not shown.}
\label{fig1}
\end{figure}

In this work, we consider the crystal structure, phonon spectra, and phase
transitions in a number of various objects in the BaS--PbS system. These include
ordered superstructures, (BaS)$_m$/(PbS)$_n$ superlattices with $m+n = 4$, 6, 8,
10, disordered solid solutions Ba$_{1-x}$Pb$_x$S with $x = 0.25$, 0.5, 0.75, and
a PbS quantum wire in a $2\times2$ BaS matrix. The investigated superstructures
include those with the metal atom ordering of the Cu$_3$Au ($L1_2$), CuAuI ($L1_0$),
and CuPt ($L1_1$) types (see Fig.~\ref{fig1}) as well as superstructures with
a metal atom ratio of 7:1 in cells with a doubled lattice parameter. All of the
considered superlattices (SLs) have the [001] orientation. To model disordered
solid solutions, we utilized special quasi-random structures SQS8 and
SQS16~\cite{PhysRevB.42.9622}.

The physical properties of the objects were calculated from first principles within
the density functional theory using the ABINIT program~\cite{abinit3}. To reduce
systematic errors in calculating the lattice parameters and the associated
physical properties, ONCPSP pseudopotentials~\cite{PhysRevB.88.085117}
constructed for the PBEsol exchange-correlation energy
functional~\cite{PhysRevLett.100.136406} were used. The maximum plane wave energy
used in the calculations was 35~Ha (952~eV). Integration over the Brillouin zone
was performed on an 8$\times$8$\times$8 Monkhorst--Pack mesh for the initial cubic
BaS and PbS and on meshes with an equivalent ${\bf k}$-point density for more
complex structures. The lattice parameters and equilibrium positions of the
atoms were found by relaxing the structure until the Hellmann--Feynman forces
reached a value
below $5 \times 10^{-6}$~Ha/Bohr (0.25~meV/{\AA}). The total energy was calcualated
with an accuracy of 10$^{-10}$~Ha. The phonon spectra were calculated
using formulas obtained from the DFPT perturbation theory. The on-site force
constants were calculated using the technique described in~\cite{PhysRevB.50.13035}.
The polarization values were calculated using the Berry phase method. The calculated
lattice parameters for cubic BaS and PbS (6.3459 and 5.9046~{\AA}) were approximately
0.6\% smaller than the experimental values 6.3877 and 5.9362~{\AA}~\cite{LB-6}.

All studied objects were dielectrics with a band gap ranging from 0.94 to 2.22~eV,
as determined in the GGA PBEsol approximation.

\section{Results and their discussion}

The structural parameters and elastic properties of the high-symmetry phases
of superstructures in the BaS--PbS system obtained in this work are in good
agreement with the previously published data~\cite{JAlloysComp.694.1348}
calculated using the FP-LAPW method and the Wu-Cohen GGA exchange-correlation
functional~\cite{PhysRevB.73.235116}.

\begin{figure*}
\centering
\includegraphics{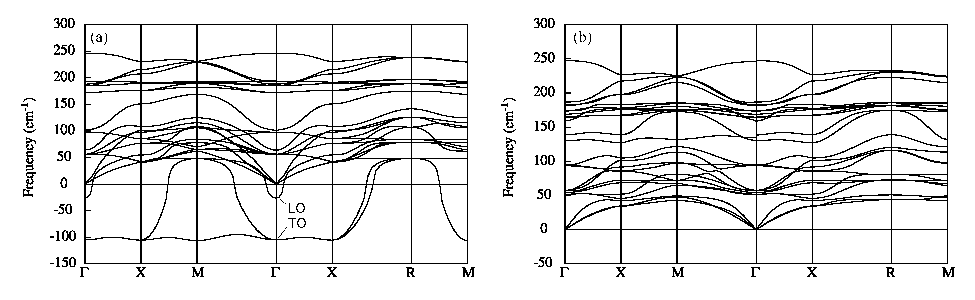}
\caption{Phonon spectrum of the Ba$_3$PbS$_4$ superstructure of the
$L1_2$ type. ($a$) data for the paraelectric $Pm{\bar 3}m$ phase, ($b$) data for
the ferroelectric $R3m$ phase. The left figure shows the LO--TO splitting of the
unstable mode at the $\Gamma$ point.}
\label{fig2}
\end{figure*}

\begin{figure*}
\centering
\includegraphics{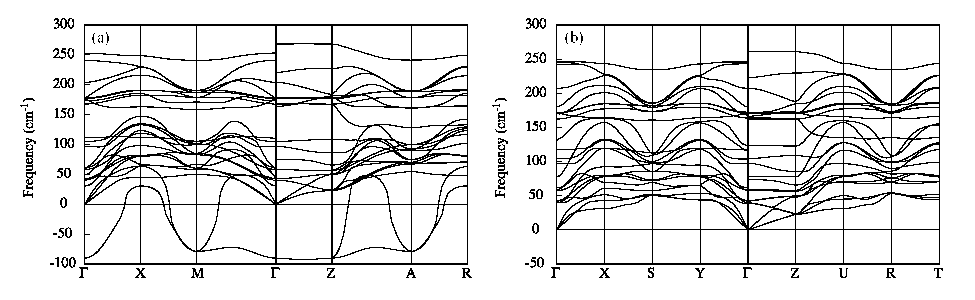}
\caption{Phonon spectrum of the (BaS)$_3$/(PbS)$_1$ superlattice
for ($a$) the paraelectric $P4/mmm$ phase, ($b$) the ferroelectric $Pmm2$ phase.}
\label{fig3}
\end{figure*}

\begin{figure}
\centering
\includegraphics{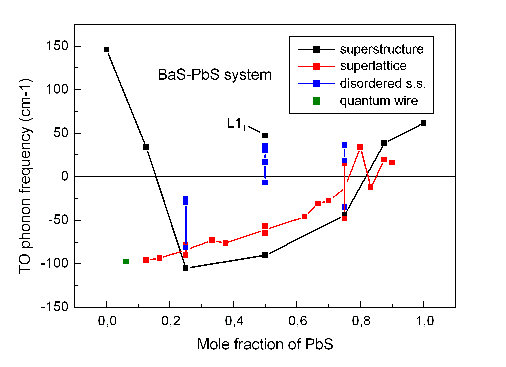}
\caption{TO phonon frequencies in superstructures, superlattices,
quantum wires, and disordered solid solutions of the BaS--PbS system as a function
of the mole fraction of PbS in their composition. The blue lines connecting the
points for solid solutions show the frequency spread for four realizations of
random disorder in the solid solutions.}
\label{fig4}
\end{figure}

\subsection{Ferroelectric instability of structures}

The following fact was found to be unexpected: calculations of the phonon spectra
for all the superstructures studied in~\cite{JAlloysComp.694.1348}, as well as for most
of the other structures listed in Sec.~\ref{sec2}, revealed a ferroelectric instability
in them (see Fig.~\ref{fig2}{\it a} and Fig.~\ref{fig3}{\it a}). The frequency of
the unstable TO phonon at the center of the Brillouin zone was found to depend on
the overall concentration of PbS%
    \footnote{As will be shown below, due to weak interaction between the
    individual components of the structure, the overall PbS concentration (the
    relative portion of PbS in the composition of the structure) is a fairly
    good parameter for characterizing the structure; finer details will only
    appear in the properties of superlattices with different layer thicknesses.}
(Fig.~\ref{fig4}). It is evident that this instability is characteristic of
all superstructures and superlattices in the range of overall PbS concentrations
of 25--75\%, as well as for the quantum wire. In solid solutions, the instability
is less pronounced. In all studied objects, the instability region is somewhat
shifted towards the region of low PbS concentrations. The only structure in which
the ferroelectric instability was absent was the CuPt ($L1_1$) superstructure with
a PbS concentration of 50\%. The reason for this anomaly will be explained below.

Calculations of the energies of all possible low-symmetry phases resulting from
the ferroelectric distortion of the paraelectric phase enable us to
determine the phase with the lowest energy and make sure that the phonon frequencies
at all points of the Brillouin zone in it are positive (Figs.~\ref{fig2}{\it b}
and \ref{fig3}{\it b})) and the elastic moduli matrix is positive-definite.
As an example, Table~\ref{table1} presents the relative energies (per molecule)
of the various ferroelectric phases for the three polar superstructures studied
in this work. It is seen that the ferroelectric ordering energy in the
superstructures systematically decreases with increasing the PbS content. The
relatively high energy of 29.45~meV for the $L1_2$ superstructure with 25\%
PbS suggests that the Curie temperature in this structure may reach 300~K.
Similar trends in the change in the energy of ferroelectric ordering are also
observed for superlattices (see Table~\ref{table2}), although the maximum gain
in the energy of ferroelectric ordering in them is smaller (8.85~meV).

\begin{table*}
\caption{\label{table1}Energies of distorted ferroelectric phases (per molecule)
and the polarization vector components for BaS--PbS superstructures. For each
superstructure, the energy of the high-symmetry phase is taken as the reference
energy.}
\begin{ruledtabular}
\begin{tabular}{ccccc}
Superstructure         & Space group   & Energy (meV) & $P_x = P_y$ (C/m$^2$) & $P_z$ (C/m$^2$) \\
\hline
Ba$_3$PbS$_4$ ($L1_2$) & $Pm{\bar 3}m$ & 0        & ---    & --- \\
                       & $P4mm$        & $-$13.41 & 0      & 0.1041 \\
                       & $Amm2$        & $-$22.70 & 0.0945 & 0 \\
                       & $R3m$         & $-$29.45 & 0.0869 & 0.0869 \\
BaPbS$_2$ ($L1_0$)     & $P4/mmm$      & 0        & ---    & --- \\
                       & $P4mm$        & $-$13.04 & 0      & 0.1926 \\
                       & $Cm$          & $-$18.75 & 0.1155 & 0.1803 \\
                       & $Pm$          & $-$20.36 & 0.1290 & 0.1765 \\
BaPb$_3$S$_4$ ($L1_2$) & $Pm{\bar 3}m$ & 0        & ---    & --- \\
                       & $P4mm$        & $-$0.87  & 0      & 0.1297 \\
                       & $Amm2$        & $-$1.44  & 0.1078 & 0 \\
                       & $R3m$         & $-$1.87  & 0.0972 & 0.0972 \\
\end{tabular}
\end{ruledtabular}
\end{table*}

\begin{table*}
\caption{\label{table2}Energies of the most stable ferroelectric phases (per molecule)
and the polarization vector components for various superlattices in the BaS--PbS system.
For each superlattice, the energy of the high-symmetry phase is taken as the
reference energy.}
\begin{ruledtabular}
\begin{tabular}{cccccc}
Superlattice   & $x$   & Space group   & Energy (meV) & $P_x = P_y$ (C/m$^2$) & $P_z$ (C/m$^2$) \\
\hline
7BaS/1PbS      & 0.125 & $P4/mmm$      & 0 \\
               &       & $Amm2$        & $-$4.45 \\
               &       & $Pmm2$        & $-$5.66 & 0.037 & 0 \\
5BaS/1PbS      & 0.167 & $P4/mmm$      & 0 \\
               &       & $Pmm2$        & $-$6.96 & 0.049 & 0 \\
3BaS/1PbS      & 0.250 & $P4/mmm$      & 0 \\
               &       & $Amm2$        & $-$6.90 \\
               &       & $Pmm2$        & $-$8.85 & 0.072 & 0 \\
6BaS/2PbS      & 0.250 & $P4/nmm$      & 0 \\
               &       & $Pmn2_1$      & $-$5.43 & 0.064 & 0 \\
4BaS/2PbS      & 0.333 & $P4/nmm$      & 0 \\
               &       & $Pmn2_1$      & $-$5.33 & 0.082 & 0 \\
5BaS/3PbS      & 0.375 & $P4/mmm$      & 0 \\
               &       & $Pmm2$        & $-$5.60 & 0.089 & 0 \\
2BaS/2PbS      & 0.500 & $P4/nmm$      & 0 \\
               &       & $Abm2$        & $-$2.18 \\
               &       & $Pmn2_1$      & $-$3.04 & 0.105 & 0 \\
3BaS/3PbS      & 0.500 & $P4/mmm$      & 0 \\
               &       & $Amm2$        & $-$4.14 \\
               &       & $Pmm2$        & $-$5.46 & 0.104 & 0 \\
3BaS/5PbS      & 0.625 & $P4/nmm$      & 0 \\
               &       & $Pmm2$        & $-$1.14 & 0.098 & 0 \\
2BaS/4PbS      & 0.667 & $P4/nmm$      & 0 \\
               &       & $Pmn2_1$      & $-$0.29 & 0.075 & 0 \\
3BaS/7PbS      & 0.700 & $P4/mmm$      & 0 \\
               &       & $Amm2$        & $-$0.128 \\
               &       & $Pmm2$        & $-$0.161 & 0.065 & 0 \\
1BaS/3PbS      & 0.750 & $P4/mmm$      & 0 \\
               &       & $Amm2$        & $-$0.037 \\
               &       & $Pmm2$        & $-$0.046 \\
               &       & $P4mm$        & $-$1.036 \\
               &       & $Cm$          & $-$1.041 \\
               &       & $Pm$          & $-$1.043 & 0.025 & 0.116 \\
1BaS/5PbS      & 0.833 & $P4/mmm$      & 0 \\
               &       & $P4mm$        & $-$0.004 & 0 & 0.031 \\
\end{tabular}
\end{ruledtabular}
\end{table*}

For obtained ferroelectric phases with the lowest energy, the magnitude
and direction of the spontaneous polarization vector were calculated. The
polarization vector is directed along the [111] axis for both cubic
superstructures of the $L1_2$ type (the $R3m$ phase), and along the
[$\xi\xi\eta$] direction for the $L1_0$ superstructure (the $Pm$ phase)
(see Table~\ref{table1}). In
superlattices (SLs) with 12.5--70\% PbS, the polarization vector lies in the
superlattice plane and is directed along the [110] axis of the pseudocubic cell
(see Table~\ref{table2}). However, in the concentration range of 75--83\% PbS,
strong fluctuations appear in the TO phonon frequencies of superlattices.
This is because the structures with the same overall PbS concentration,
but with different layer thicknesses, are compared in one figure. In this
concentration range, the instability of TO phonons polarized in the $z$~direction
is observed in SLs containing one BaS monolayer, such as (BaS)$_1$/(PbS)$_3$
and (BaS)$_1$/(PbS)$_5$ SLs. However, this instability disappears in (BaS)$_1$/(PbS)$_7$
and (BaS)$_1$/(PbS)$_9$ SLs. In SLs containing two BaS monolayers, such as
(BaS)$_2$/(PbS)$_6$ and (BaS)$_2$/(PbS)$_8$, this instability is absent.
Due to the rapid weakening of the ferroelectric distortion in the
$xy$ plane, the instability in the $z$ direction causes the polarization vector
to turn towards the $z$ axis direction. As the PbS concentration increases, the
polarization in SLs first increases, reaching a maximum of approximately 0.15~C/m$^2$
at 50\% PbS, and then decreases, with the vector changing its direction at
$x \ge 0.75$ (see Table~\ref{table2}). In polar solid solutions, the direction
of polarization depends on the used atomic configuration and is often close to the
[110] direction of the pseudocubic cell, and less often close to the [100]
direction.

\subsection{Search for the ground state structure}

An analysis of the phonon spectra of high-symmetry phases reveals the presence
of instabilities at points other than the $\Gamma$ point of the Brillouin zone
(see Figs.~\ref{fig2}{\it a} and \ref{fig3}{\it a}). These phonon spectra
can be explained by the chain-structure instability model, which was first
discovered in KNbO$_3$~\cite{PhysRevLett.74.4067} and later observed in
BaTiO$_3$~\cite{Ferroelectrics.206.205} and KNbO$_3$/KTaO$_3$~\cite{PhysStatusSolidiB.249.789}
and BaTiO$_3$/BaZrO$_3$ superlattices~\cite{PhysSolidState.55.1198}. The authors of the
papers~\cite{SolidStateCommun.7.589,JETP.115.309} also came to the conclusion
about the chain-structure instability in BaTiO$_3$ and KNbO$_3$ crystals when
explaining the unusual X-ray scattering patterns observed in experiments.

Indeed, an analysis of the character of atomic displacements in the eigenvectors
of unstable modes reveals that the displacements occur in infinite --Pb--S--Pb--S--
chains propagating along the fourfold axes of the cubic NaCl structure. These
eigenvectors describe out-of-phase displacements of Pb and S atoms along the chain,
but the resulting polarization in the chain can be either parallel or antiparallel
to the polarization in adjacent chains. The absence of a ferroelectric
instability in the CuPt-type superstructure (black dot in Fig.~\ref{fig4}) is
consistent with this explanation, as infinite --Pb--S--Pb--S-- chains cannot exist
in this structure. The conclusion about the chain-structure instability is further
supported by the data obtained for an isolated infinite chain of Pb--S atoms
embedded in a BaS matrix, known as a quantum wire (green dot in Fig.~\ref{fig4}).
The observed instability of phonons in the $z$ direction in the (BaS)$_1$/(PbS)$_3$
superlattice with an overall PbS concentration of 75\% can be attributed to the
fact that the presence of one BaS monolayer breaks only one of the two infinite
--Pb--S--Pb--S-- chains in the $z$ direction, while the other chain remains intact.
To completely break these chains, two or more BaS monolayers are required.

\begin{table*}
\caption{\label{table3}Energies of phases and the character of their stability for
the Ba$_3$PbS$_4$ ($L1_2$) superstructure.}
\begin{threeparttable}
\begin{ruledtabular}
\begin{tabular}{cccc}
Original phase and &          &               & Stability/ \\
unstable phonon    & Phase    & Energy (meV)  & the relaxation path \\
\hline
$\Gamma$       & $Pm{\bar 3}m$ & 0   & relaxes to $P4mm$, $Amm2$, $R3m$ \\
               & $P4mm$   & $-$13.41 & relaxes to $Amm2$, $R3m$; $Pma2$ \\
               & $Amm2$   & $-$22.70 & relaxes to $R3m$; $Pc$, $Pmn2_1$ \\
               & $R3m$    & $-$29.45 & metastable \\
$X(\eta,0)$    & $Pmma$   & $-$14.03 & relaxes to $Pmc2_1$, $Pma2$; $P2_1/m$ \\
$X(\eta,\eta)$ & $Cmcm$   & $-$23.14 & relaxes to $Cmc2_1$; $P2_1/c$, $Pnma$ \\
$M$            & $P4/nmm$ & $-$13.83 & relaxes to $Pmn2_1$, $Abm2$ \\
\hline
$Pmma$+polar($\eta,0$) & $Pmc2_1$ & $-$23.23 & relaxes to $Cmc2_1$, $Pc$ \\
$Pmma$+nonpolar        & $P2_1/m$ & $-$23.14 & relaxes to $Cmc2_1$ \\
$Pmma$+polar($0,\eta$)\tnote{a} & $Pma2$   & $-$23.08 & relaxes to $Pc$ \\
$Cmcm$+polar           & $Cmc2_1$ & {\bf $-$29.85} & stable (the ground state) \\
$P4/nmm$+polar($\eta,0$) & $Abm2$ & $-$23.01   & relaxes to $Pmn2_1$ \\
$P4/nmm$+polar($\eta,\eta$)\tnote{a} & $Pmn2_1$ & $-$29.72 & metastable \\
$Cmcm$+$X$-phonon & $P2_1/c$ & $-$29.62 & metastable \\
$Cmcm$+$S$-phonon & $Pnma$   & $-$29.44 & metastable \\
\hline
$Pmc2_1$+polar\tnote{a} & $Pc$ & $-$29.79 & metastable \\
\end{tabular}
\end{ruledtabular}
\begin{tablenotes}
 \item[a] One of the relaxation paths to the specified phase is indicated.
\end{tablenotes}
\end{threeparttable}
\end{table*}

We will now examine the Ba$_3$PbS$_4$ superstructure ($L1_2$ type) in more detail.
Its structure, shown in Fig.~\ref{fig1}, is characterized by the presence of three
infinite Pb--S chains propagating along the fourfold axes. Calculations of the
phonon spectrum for this superstructure reveal instabilities at the $\Gamma$,
$X$, and $M$ points (see Fig.~\ref{fig2}{\it a}). The ferroelectric instability
at the $\Gamma$ point results in the well-known tetragonal, orthorhombic, and
rhombohedral phases (Table~\ref{table3}). The energies of these phases, resulting
from the condensation of all unstable phonons in the phonon spectrum in this
superstructure, are presented in Table~\ref{table3}. In the obtained data,
three groups of the energies are clearly distinguished (13--14~meV, 22--23~meV,
and 29~meV). The analysis of the structure of these phases reveals that they
correspond to structures in which the polarization is established in one, two,
and three infinite chains, respectively.

\begin{figure}
\centering
\includegraphics[scale=0.4]{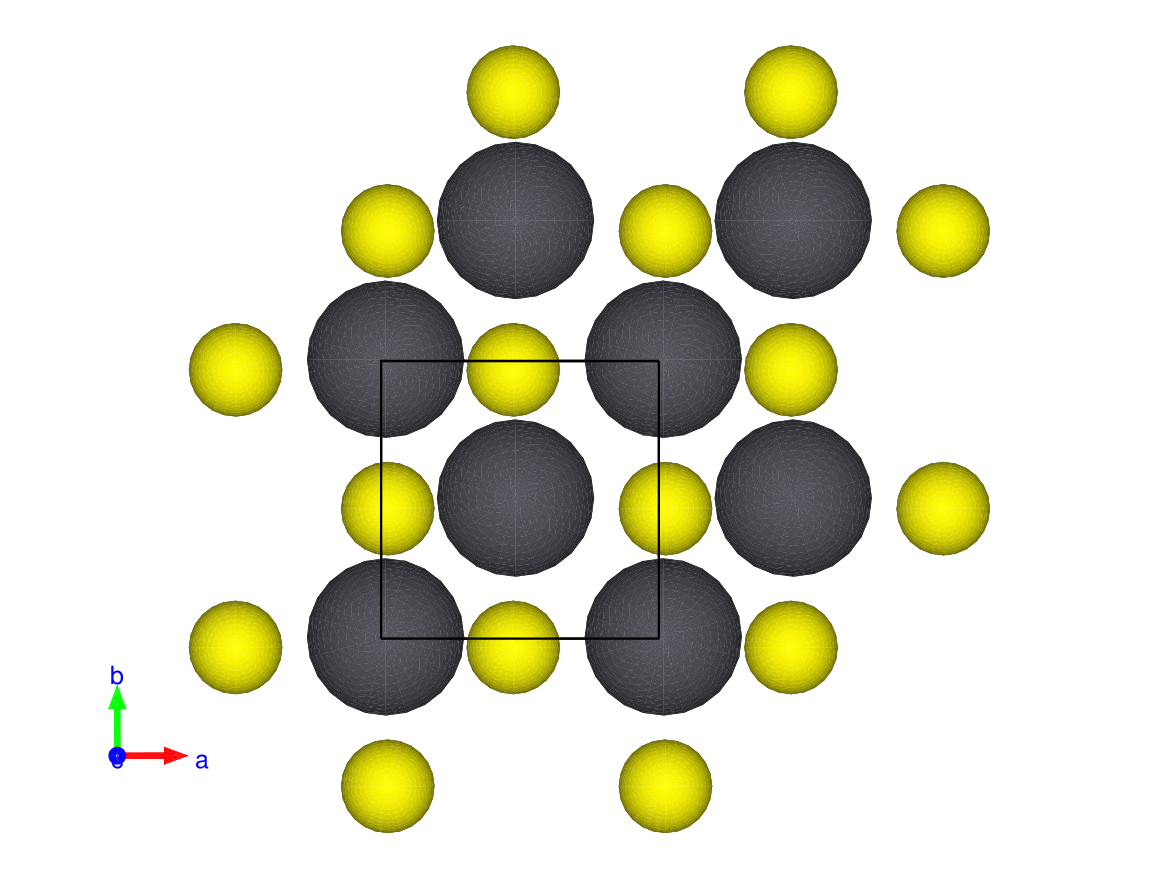}
\caption{(001) section of the ground state structure for the (BaS)$_2$/(PbS)$_2$
superlattice ($Pca2_1$ space group). The picture shows parallel polar displacements
in two chains along the $y$~axis and opposite polar displacements in two chains
along the $x$~axis, which give a zero net polarization in the $x$~direction. The
rectangle shows the unit cell, the Pb atoms are gray, and the S atoms are yellow.}
\label{fig5}
\end{figure}

An instability at points on the boundary of the Brillouin zone results in the
doubling of the unit cell and the emergence of one or two directions in which
adjacent chains are polarized out of phase (see Fig.~\ref{fig5}). This results
in the net
polarization of zero in those directions. A check for stability of the obtained phases
(stable structures should have positive frequencies of all optical phonons
and a positive definiteness of the elastic moduli matrix) shows that phases with
energies of 13--14~meV and 22--23~meV remain unstable at least at the $\Gamma$
point. They indicate the ferroelectric instability of the chains that remained
unpolarized. Continuing the search for the ground state structure and relying on
the results of calculations of the phonon spectra in the obtained structures, we
can go through a chain of phase transformations until a structure that meets
the stability criteria is found. In the phases with the lowest energies, all
chains are found to be always polarized. However, if we take into account that
instabilities can be also detected at the Brillouin zone boundary in some
intermediate structures, then using this algorithm we can even find phases with
zero polarization, despite all chains being polarized. This type of ordering
can be logically referred to as antiferroelectric (AFE). We conclude that
a superstructure under consideration can contain not only a ferroelectric (FE)
phase with maximum polarization, but also phases with reduced or even zero
polarization. The stable phases with reduced polarization contain pairs of
chains with opposite polarizations, in addition to singly polarized chains.
These phases are referred to as phases with mixed (FE+AFE) ordering. All
solutions corresponding to FE, AFE, and mixed ordering are stable, indicating that
spontaneous reorientation of the polarization direction does not occur in them.

\begin{table}
\caption{\label{table4}Polarization in different low-energy phases of the
Ba$_3$PbS$_4$ ($L1_2$) superstructure.}
\begin{ruledtabular}
\begin{tabular}{ccc}
Phase    & Energy (meV) & Polarization (C/m$^2$) \\
\hline
$Pnma$   & $-$29.44 & (0.0, 0.0, 0.0) \\
$R3m$    & $-$29.45 & (0.0869, 0.0869, 0.0869) \\
$P2_1/c$ & $-$29.62 & (0.0, 0.0, 0.0) \\
$Pmn2_1$ & $-$29.72 & (0.0874, 0.0874, 0.0) \\
$Pc$     & $-$29.79 & (0.0, 0.0870, 0.0874) \\
$Cmc2_1$ & {\bf $-$29.85} & (0.0, 0.0, 0.0875) \\
\end{tabular}
\end{ruledtabular}
\end{table}

The results of polarization calculations for the low-energy phases of the
Ba$_3$PbS$_4$ superstructure are presented in Table~\ref{table4}. The energies
of the six phases in the table differ by less than 0.5~meV, but their
polarization values vary significantly. However, it is worth noting that the
polarization values themselves, if nonzero, remain almost identical in each
direction. As mentioned earlier, a zero polarization in any direction indicates
that the unit cell contains two chains with opposite polarization. It is
interesting to note that among the considered structures, the mixed FE+AFE
$Cmc2_1$ phase has the lowest energy and is therefore the ground state structure.

The small energy difference between structures with parallel and antiparallel
orientation of polarization in adjacent chains, which is 1 to 2 orders of magnitude
smaller than the ferroelectric ordering energy, indicates weak interactions
between the adjacent chains. This is likely the reason for the emergence of
many metastable states in the superstructure and for a quite accurate
characterization of their properties by the overall PbS concentration.

In the BaPbS$_2$ ($L1_0$) superstructure, the number of infinite chains doubles
due to the doubling of the number of Pb atoms. This results in a significant
increase in the number of possible polarization configurations in the chains, with
a total of approximately 50 possible phases. A table of these configurations is
provided in the Supplementary material; however, for brevity, we will only
mention the ground state structure here. This structure is described by the $P2_1$
space group and belongs to structures with mixed FE+AFE ordering. The energy gain
upon transition to this structure is 24.12~meV, and the polarization is equal to
0.148~C/m$^2$.

\begin{table*}
\caption{\label{table5}Energies of phases and the character of their stability
for the (BaS)$_3$/(PbS)$_1$ superlattice.}
\begin{threeparttable}
\begin{ruledtabular}
\begin{tabular}{cccc}
Original phase and &          &              & Stability/ \\
unstable phonon    & Phase    & Energy (meV) & the relaxation path \\
\hline
$\Gamma$       & $P4/mmm$ & 0        & relaxes to $Amm2$, $Pmm2$ \\
               & $Amm2$   & $-$6.90  & relaxes to $Pmm2$; $Ama2$, $Pmc2_1$ \\
               & $Pmm2$   & $-$8.85  & metastable \\
$M(\eta,0)$    & $Pmma(2)$ & $-$8.37 & relaxes to $Pmc2_1$, $Cmmm$ \\
$M(\eta,\eta)$ & $Cmmm$   & $-$11.34 & metastable \\
$Z(\eta,0)$    & $Pmma$   & $-$8.86  & metastable \\
$Z(\eta,\eta)$ & $Cmcm$   & $-$6.91  & relaxes to $Pmma$, $Ama2$; $Pnma$, $Pbcm$
\\
$A(\eta,0)$    & $Imma$   & $-$8.32  & relaxes to $Ima2$, $C2/m$ \\
$A(\eta,\eta)$ & $Fmmm$   & $-$11.30 & metastable \\
\hline
$Pmma$(2)+polar\tnote{a} & $Pmc2_1$ & $-$13.752 & metastable \\
$Pmma$(2)+phonon1@$X$ & $Pbam$    & $-$13.34 & metastable \\
$Pmma$(2)+phonon2@$X$ & $Pbam$(2) & $-$10.15 & metastable \\
$Cmcm$+polar\tnote{a}  & $Ama2$   & $-$8.44  & metastable \\
$Cmcm$+phonon1@$S$    & $Pnma$ & {\bf $-$13.773} & stable (the ground state) \\
$Cmcm$+phonon2@$S$    & $Pbcm$ & $-$13.757 & metastable \\
$Imma$+polar    & $Ima2$   & $-$13.766 & metastable \\
$Imma$+nonpolar & $C2/m$   & $-$11.31  & metastable \\
\end{tabular}
\end{ruledtabular}
\begin{tablenotes}
 \item[a] One of the relaxation paths to the specified phase is indicated.
\end{tablenotes}
\end{threeparttable}
\end{table*}

As another example, let us consider the properties of the (BaS)$_3$/(PbS)$_1$
superlattice. The structure of this superlattice (see Fig.~\ref{fig1}) is characterized
by the suppression of the formation of two infinite Pb--S chains along the~$z$ axis,
as the Pb atoms are ``isolated'' from each other by BaS layers. However, two
infinite --Pb--S--Pb--S-- chains remain in the structure in the $x$ and $y$~directions.
The primitive cell of this superlattice, in the $P4/mmm$ phase, is constructed using
the vectors (1/2,1/2,0) and ($-$1/2,1/2,0) of the basal plane and contains one
Pb atom. Therefore, the number of possible structures in this system is not very
large (see Table~\ref{table5}). Out of these structures, only four are stable
ferroelectric phases. In a cell with one Pb atom, the only possible phase with
the polarization in both chains is the $Pmm2$ phase, with the polarization
vector along the [110] direction of the pseudocubic cell.
In the three other polar phases that arise from the condensation of phonons
at the boundary of the Brillouin zone, the unit cells contain two Pb atoms each,
doubling the number of chains. However, since the polarization directions must be
opposite in at least two chains (this is dictated by the condensation wave
vector), either the ferroelectric phases with [100] ($Ama2$) or [010] ($Pmc2_1$,
$Ima2$) polarizations emerge, or the polarization in the cell is exactly zero.
All of these phases are stable, and all chains within them are polarized. However,
the amplitudes of the polar displacements in the chains vary greatly among the
different phases, resulting in a large spread in their energies.
Calculations have shown that the ground state in the (BaS)$_3$/(PbS)$_1$
superlattice is the AFE phase $Pnma$, with all other ferroelectric phases being
metastable.

In this work, calculations for more complex superstructures and superlattices
were also performed. They fully confirm the conclusion about chain-structure
ferroelectric instability in these structures. However, the number of potential
phases that must be analyzed is very large. The results for some of these
structures can be found in the Supplementary material.

Our calculations indicate that a purely ferroelectric ground state is not
very common for the studied structures. It is present in the BaPb$_3$S$_4$
superstructure (in the $R3m$ phase) as well as in the (BaS)$_2$/(PbS)$_4$
($Pmn2_1$) and (BaS)$_1$/(PbS)$_3$ ($Cm$) superstructures.%
    \footnote{The latter structure can be more accurately described as
    \emph{ferri}electric, as it contains two polar fragments with opposite
    polarization along the $z$ axis that are not precisely compensated.}
The Ba$_3$PbS$_4$ ($Cmc2_1$) and BaPbS$_2$ ($P2_1$) superstructures, as well as
the (BaS)$_2$/(PbS)$_2$ ($Pca2_1$) superlattice, exhibit a mixed FE+AFE ground
state. The AFE ground state is observed in the (BaS)$_3$/(PbS)$_1$ ($Pnma$),
(BaS)$_4$/(PbS)$_2$ ($P{\bar 4}2m$), and (BaS)$_3$/(PbS)$_3$ ($Pmma$)
superlattices. It is clear that the mixed or antiferroelectric ground state
is more common for the studied structures. This is likely due to the negative
sign of interaction between adjacent chains. For example, in Ba$_3$PbS$_4$,
the energy differences between the $Pmma/P4mm$ and $Cmcm/Amm2$ pairs of phases
with parallel and antiparallel orientations of polarization in the chains are
$-$0.62 and $-$0.44~meV, respectively (see Table~\ref{table3}). Similarly, in
the (BaS)$_3$/(PbS)$_1$ superlattice, the energy differences between the
$Pmma(2)/Amm2$ and $Cmmm/Pmm2$ pairs of phases with similar chains are $-$1.47
and $-$2.49~meV, respectively (see Table~\ref{table5}).

The obtained phonon spectra show that unstable phonons are also observed in
the discussed structures for a wide range of intermediate values of the
wave vector~${\bf q}$. For instance, in the Ba$_3$PbS$_4$ superstructure
(see Fig.~\ref{fig2}(a)), they are observed over the entire plane with
$q_z = 0$ . The eigenvectors of these phonons consist of chains of alternating
transverse polar displacements of atoms. Upon relaxation, these phonons result
in the formation of structures with a total polarization of zero and similar
phase energies. For example, the energy of a phase formed from the condensation
of a doubly degenerate mode at the point with ${\bf q} = (0.0, 0.0, 0.25)$ and
satisfying all the stability criteria is $-$29.73~meV, which is only 0.12~meV
higher than the energy of the ground state found for this superstructure
(see Table~\ref{table3}). Since each ${\bf q}$ value has its own unique spatial
configuration of polarized chains, the existence of a very dense mesh of wave
vectors implies the emergence of a virtually infinite number of minima in the
configuration space, separated by potential barriers.

The obtained results suggest the emergence of a complex multi-minimum potential
with nearly degenerate energy states separated by potential barriers in the
configuration space of the systems under consideration. As a result, nonergodicity
in their physical properties can be expected.
This is why the task of finding the ground states in all studied structures, as
posed in this work, is practically unsolvable. The medium that emerges in these
systems is suitable for recording information in the form of polarization,
with its magnitude determined by the fraction of chains polarized in a given
direction. Alternatively, the obtained medium can be seen as an ultrafine domain
structure of a ferroelectric. One can expect that the considered structures can
be switched between different phases by an external electric field, resulting
in phenomena similar to the \emph{irreversible} transition from the AFE to the
FE phase observed in NaNbO$_3$~\cite{ActaMater.200.127}. At low temperatures,
the response to an external electric field is likely to exhibit high ``viscosity''
(i.e. relaxor behavior).

\subsection{The origin of the ferroelectric instability}

A comparison of the phonon spectra of the above-considered structures with those
of bulk oxide perovskites KNbO$_3$~\cite{PhysRevLett.74.4067}, BaTiO$_3$,
and RaTiO$_3$~\cite{PhysSolidState.51.362} reveals their qualitative similarity.
This suggests that the mechanism of the ferroelectric instability in
these materials is the same~--- the chain-structure instability. However, unlike
other perovskites, the structures under consideration lack the antiferrodistortive
instability associated with octahedral rotations in perovskites. Instead, they
may exhibit antiferroelectric structures with opposite polarization orientations
in adjacent chains. Calculations show that the character of the interaction between
chains in these two systems differs significantly. In BaTiO$_3$, the energy
difference between the ferro- and antiferroelectric phases of $P4mm$ and $Pmma$
is +2.20~meV, while in KNbO$_3$ it is +5.42~meV, indicating that the parallel
orientation of polarization in adjacent chains is energetically more favorable.
However, as mentioned earlier, in Ba$_3$PbS$_4$ and the (BaS)$_3$(PbS)$_1$ superlattice, the energy
difference is negative, suggesting that the antiparallel orientation of polarization
in adjacent chains is more favorable. In perovskites, a sufficiently strong
interaction between chains results in bulk ferroelectricity, whereas in the
systems under consideration, the interaction is frustrated and weak, resulting
in the emergence of complex polar structures. This is why the ground
state in these systems is not purely ferroelectric. Nevertheless, a common
property of the structures emerging in the systems under consideration is that
they are composed of one-dimensional ferroelectrically polarized chains, which
may be randomly oriented in the perpendicular direction.

To investigate the relationship between the emerging ferroelectric phenomena in the BaS--PbS system and
the potential off-centering of Pb atoms, which may be caused by the presence of
a lone $s^2$ electron pair in Pb$^{2+}$ ions, the on-site force constant for the
Pb atom in a superstructure with 12.5\% PbS was calculated. The calculations
yielded a value of +0.02290~Ha/Bohr$^2$ for this constant, which was obtained
by averaging over 8 ${\bf q}$-points \cite{PhysRevB.50.13035}. Its positive value
indicates that when the Pb atom is displaced from its site, the restoring force
is directed towards the site, suggesting the absence of off-centering. Therefore,
the ferroelectric instability in the BaS--PbS system is likely caused by strongly
correlated collective displacements of atoms in the chains, rather than by the
off-centering of Pb. The high effective charge of the Pb$^{2+}$ ions ($Z^* = 3.3$--4.3)
further supports the significant role of the electron density redistribution in
the Pb--S covalent bonds upon atomic displacement and its contribution to
establishing long-range order.

\begin{figure}
\centering
\includegraphics{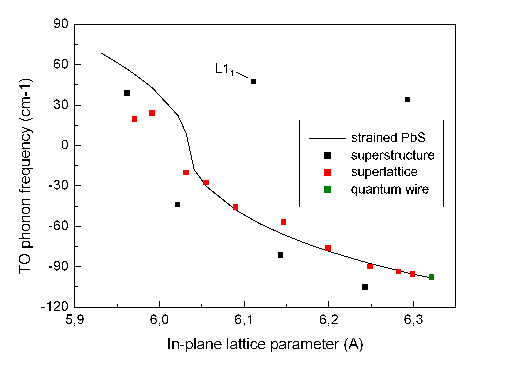}
\caption{TO phonon frequency as a function of the lattice parameter~$a$
in the studied structures (dots). For comparison, the solid curve shows the
TO$_{xy}$ phonon frequency in biaxially strained bulk PbS.}
\label{fig6}
\end{figure}

To better understand the reasons for the emergence of ferroelectricity in the
systems under consideration, it was interesting to examine the role of strain
induced by barium atoms whose ionic radius is larger than that of lead. Fig.~\ref{fig6}
shows the relationship between the frequency of the ferroelectric soft mode and the
in-plane lattice parameter in the structures being considered. For comparison, the
figure also includes the dependence of the TO$_{xy}$ phonon frequency in bulk
PbS under biaxial stretching. It is evident that both the samples under consideration
and bulk PbS exhibit a similar trend of increased ferroelectric instability under
stretching strain. This suggests that the ferroelectric behavior observed in the
BaS--PbS system is a result of the stretching effect of the BaS matrix on the
ferroelectric instability in the Pb--S--Pb--S chains.

As we have seen, the ferroelectric instability is significantly weaker in
disordered solid solutions of the BaS--PbS system. This can be attributed to
the fact that the formation of infinite --Pb--S--Pb--S-- chains is highly
unlikely in such structures. To determine the minimum length of the chain
necessary for local instability to occur, additional calculations were performed.
These calculations revealed that for the ferroelectric instability to occur, the
chain must consist of at least three Pb atoms arranged in a linear fashion.

\begin{table}
\caption{\label{table6}The bulk modulus (in GPa) for different superstructures
in the BaS--PbS system.}
\begin{ruledtabular}
\begin{tabular}{cccc}
Composition           & Data of                     & \multicolumn{2}{c}{This work} \\
\cline{3-4}
(mole fraction of Pb) & \cite{JAlloysComp.694.1348} & paraphase & ground state \\
\hline
0 (BaS)       & 44.40 & 46.14 & --- \\
0.125         & ---   & 47.55 & --- \\
0.25 ($L1_2$) & 45.23 & 48.86 & 39.65 \\
0.5 ($L1_0$)  & 49.44 & 52.15 & 33.04 \\
0.5 ($L1_1$)  & ---   & 52.58 & --- \\
0.75 ($L1_2$) & 54.02 & 55.92 & 41.56 \\
0.875         & ---   & 58.00 & --- \\
1.0 (PbS)     & 59.37 & 60.12 & --- \\
\end{tabular}
\end{ruledtabular}
\end{table}

\subsection{Other physical properties}

In most of the considered structures, the paraelectric phase was found to be
ferroelectrically distorted. Therefore, it was necessary to recalculate the
values of the elastic moduli, which had previously been determined only for the
high-symmetry phase. The bulk modulus was calculated using the Reuss scheme~\cite{ProcPhysSocA.65.349}
from the elastic moduli matrix. The results of these calculations are given in
Table~\ref{table6}. The values for the high-symmetry phases are in good agreement
with previously published data~\cite{JAlloysComp.694.1348}; however, the presence
of structural distortions results in a decrease in the modulus by 13--37\% compared
to the high-symmetry phase. In disordered solid solutions, the bulk modulus is
also lower than in the high-symmetry phase of superstructures. For example, in
the Ba$_{0.5}$Pb$_{0.5}$S solid solution, it decreases to 50.46--50.90~GPa for
non-polar phases, and for polar phases, it is even lower (46.93--47.29~GPa).

The piezoelectric properties of the considered structures are relatively modest.
The maximum values of $d_{ijk}$ correspond to the moduli for strain along the
polar axis, with values of 6~pC/N for the ground state of the Ba$_3$PbS$_4$
superstructure, 27~pC/N for the ground state of the  BaPbS$_2$ superstructure,
and 29~pC/N for the ground state of the (BaS)$_2$(PbS)$_2$ superlattice.

\subsection{Concluding remarks}

Our calculations did not reveal any tendency for the BaS--PbS system to order
into Cu$_3$Au and CuAuI type superstructures, as was observed in the SrS--PbS
system~\cite{JAmChemSoc.136.1628} (our calculations confirmed the results
of this previous study).

The ferroelectric instability was also revealed in the selenide and telluride
analogs of the considered system. In the Ba$_3$PbSe$_4$ and Ba$_3$PbTe$_4$
superstructures, the energy gain due to the ferroelectric distortion was found
to be 20.84 and 18.00~meV per molecule, respectively. However, for the
(Ba{\it X})$_3$(Pb{\it X})$_1$ superlattices, the gain was smaller: 6.27~meV
({\it X} = Se) and 6.07~meV ({\it X} = Te). The on-site force constants of
the Pb atom were calculated to be
+0.02046 and +0.01716~Ha/Bohr$^2$ in the selenide and telluride systems,
respectively. This suggests that Pb$^{2+}$ is not an off-center ion, and therefore
the emergence of ferroelectricity in these systems is also associated with the
chain-structure instability. However, despite their similar composition,
the ferroelectricity does not emerge in superstructures and superlattices of
the SrS--PbS and CaS--PbS systems.

The calculated values of the mixing enthalpy in the Ba$_{1-x}$Pb$_x$S solid
solution range from 10 to 24~meV per molecule, indicating that the preparation
of the samples in the BaS--PbS system should not be problematic from a
thermodynamic standpoint. The mixing enthalpies for the Ba$_{0.5}$Pb$_{0.5}$Se
and Ba$_{0.5}$Pb$_{0.5}$Te solid solutions are significantly higher: 32--40~meV
and 57--67~meV per molecule, respectively.

\section{Conclusion}

A new class of materials with ferroelectric properties, including superstructures,
superlattices, quantum wires, and disordered solid solutions in the Ba$X$--Pb$X$
systems ($X = {}$S, Se, Te) with the NaCl structure, has been discovered through
first-principles calculations within the density functional theory. These
phenomena are attributed to the chain-structure ferroelectric instability
of the linear --Pb--$X$--Pb--$X$-- chains in the NaCl structure. The study has
also revealed the presence of competing antiferroelectric configurations in the
BaS--PbS system, with the ground state of the studied systems often being
antiferroelectric phases or phases with a mixed ferroelectric--antiferroelectric
(FE-AFE) state. The close proximity of the energies of ferroelectric,
antiferroelectric, and mixed FE+AFE configurations suggests the emergence of
a multi-minimum potential with an infinite number of wells separated by
potential barriers in the configuration space. This may lead to nonergodicity
in the systems at low temperatures.



\providecommand{\BIBYu}{Yu}

\end{document}